# The Impact of Large Language Models on Task Automation in Manufacturing Services

Jochen Wulf[a]*, Jürg Meierhofer[a]

[a]Zurich University of Applied Sciences (ZHAW), Technikumstrasse 81, CH-8401 Winterthur

* Corresponding author. Tel.: +41 58 934 62 27. E-mail address: jochen.wulf@zhaw.ch

**Abstract**

This paper explores the potential of large language models (LLMs) for task automation in the provision of technical services in the production machinery sector. By focusing on text correction, summarization, and question answering, the study demonstrates how LLMs can enhance operational efficiency and customer support quality. Through prototyping and the analysis of real-life customer data, LLMs are shown to reliably correct errors, generate concise summaries of complex communication, and provide accurate, context-aware responses to customer inquiries. The research also integrates Retrieval Augmented Generation (RAG) to combine LLM outputs with domain-specific knowledge, enhancing precision and relevance. While the findings highlight significant efficiency gains, challenges such as knowledge hallucination and integration with human workflows remain barriers to large-scale adoption. This study contributes to the theoretical understanding and practical application of LLMs in manufacturing, paving the way for further research into scalable, domain-specific implementations.

## 1. Introduction

The advent of Generative AI (GenAI) and Large Language Models (LLMs) represents a transformative shift in the manufacturing industry, promising substantial economic potential and operational efficiencies. Generative AI, built on foundation models capable of processing vast amounts of unstructured data, offers unprecedented capabilities in automating and optimizing various manufacturing processes. This technological advancement is poised to revolutionize the industry by enhancing productivity, reducing costs, and fostering innovation [1,2].

GenAI holds significant potential for various applications in manufacturing, from enhancing data insights by analyzing unstructured data to optimizing materials processing and mechanical design [3,4]. Despite this potential, the adoption and scaling of GenAI in the manufacturing sector face significant challenges. A recent survey conducted by McKinsey & Company revealed that while nearly 80% of large companies have begun experimenting with GenAI since the launch of ChatGPT in November 2022, only 3% have successfully scaled at least one application across their operations [5]. Approximately 24% of companies have initiated pilot projects, while 63% are still in the early stages of experimentation or have not yet started any Generative AI initiatives. This indicates that the majority of companies are still grappling with how to effectively integrate and scale these technologies within their operations.

Recent academic studies have explored the potential applications of LLMs in the manufacturing sector, focusing on product design, production planning and execution, production assistance, and product maintenance [6–13]. Despite these advances, there is still insufficient research into which manual activities can be fully automated with LLMs.

This research addresses this gap and uses a prototyping approach with real customer data to validate the automation potential of three cognitive tasks: text correction, summarization and question answering. The results contribute to the body of knowledge on the design and application of LLMs for manufacturing services and demonstrate LLMs can enhance efficiency and service quality.

## 2. Theoretical Background

Large language models have recently demonstrated remarkable capabilities in natural language processing tasks and beyond [14–16]. These models combine various technical innovations, especially in the domains of architectural design, training strategies, improvements in context length, fine-tuning, multimodality, and scope of the training datasets [17]. However, LLMs face challenges such as hallucination, outdated knowledge, and non-transparent, untraceable processing processes [18].

Retrieval Augmented Generation (RAG) has emerged as a promising solution by incorporating knowledge from external databases [19]. This improves the accuracy and credibility of the models, especially for knowledge-intensive tasks, and allows for continuous knowledge updates and the integration of domain-specific information. RAG synergistically combines the intrinsic knowledge of LLMs with the extensive, dynamic information in external databases.

Recent academic studies have started to investigate the potential applications of LLMs in the manufacturing sector, focusing primarily on production planning and execution, product maintenance, production assistance, and product design. For production planning and execution, Xia et al. [6] propose a framework that combines LLMs with a digital twin and an industrial automation system. LLM agents can access descriptive data from the digital twin system and execute production steps within the industrial automation system via service interfaces. They demonstrate the application of LLM agents in production planning and the control of flexible production for the use case of producing a steel sheet with a hole. The agent-based system retrieves the workpiece from storage, inspects it, and then drills the hole using a CNC machine. The transport robot moves the workpiece between modules, ensuring each step is completed efficiently. Similarly, Fan et al. [7] showcase the use of LLM capabilities in natural language understanding, planning, and reasoning for autonomous industrial robotics. They validate their designed system with experiments covering several multi-step manufacturing procedures, such as the coating and stacking of cubes. Both, Xia et al. [6] and Fan et al. [7] showcase the ability of LLMs to transform natural language task descriptions into executable software code, such as API calls to MES modules.

A next field of application is machine maintenance. Wang et al. [8] present a method for customizing GPT-3.5 with domain-specific knowledge for intelligent aircraft maintenance, resulting in consistent maintenance recommendations. Vidyaratne et al. [9] investigate the use of LLMs for extracting troubleshooting trees for industrial equipment, demonstrating their approach with three different product manuals and achieving high data extraction coverage and precision.

LLMs also have potential in production assistance. For instance, Freire et al. [10] introduce a system based on an LLM that retrieves information from extensive factory documentation and operator expertise. This system is designed to efficiently respond to operator queries and facilitate the dissemination of new knowledge in textile and detergent production. Colabianchi et al. [11] employ an LLM-based digital intelligent assistant in a laboratory experiment to guide operators through the box assembly process, demonstrating significant improvements in operator experience, reduced cognitive load, and enhanced output quality.

Lastly, LLMs also have potential for product design, Makatura et al. [12] examine the integration of LLMs throughout the design and manufacturing process. They illustrate how LLMs can convert text prompts into designs, generate design spaces and variations, translate designs into manufacturing instructions, assess design performance, and search for designs based on performance metrics. They demonstrate that reliability in LLM-generated output is still a challenge for fully automated systems, as initial designs frequently contain flaws and necessitate corrective human intervention. Additionally, Yang et al. [13] highlight the role of LLMs in supporting industrial data-centric R&D cycles, particularly in the context of quantitative investment research.

In summary, a broad spectrum of LLM applications have been discussed based on experiments and proof of concepts. It remains unclear, however, which types of manual activities can be automated with LLMs. This research discusses several cognitive tasks and demonstrates how to use LLMs for the provision of technical service for manufacturers.

## 3. Research Methodology

### 3.1. Theoretical Framework

An examination of the fundamental research literature on LLMs suggests five cognitive tasks addressed by LLMs relevant for application in technical service [20] (see Table 1).

Table 1. Cognitive Tasks with Potential for LLM Automation [20]

| Task | Description |
| --- | --- |
| Translation and Correction | LLMs correct and translate text between different languages. |
| Summarization | LLMs produce summaries that convey the overall meaning of the original text by understanding inherent relationships. |
| Content Generation | LLMs create diverse content types, such as emails, social media posts, blog articles, and stories. |
| Question Answering | LLMs answer questions by using factual knowledge from their training data or external contextual information provided in the prompt. |
| Reasoning | LLMs use evidence and logic to solve complex problems, involving sequential reasoning based on factual knowledge to draw conclusions. |

The first cognitive task is translation and correction. LLMs are capable of translating text between different languages or language modes by recognizing and learning the patterns and structures inherent in various languages from their training data. Furthermore, LLMs are considered an important tool for correcting errors in natural language text [21]. However, managing specialized domain vocabulary, such as technology-specific technical terms, continues to pose a challenge [22]. The second task is summarization. LLMs can grasp the relationships between different sections of a text and produce summaries that accurately convey the overall meaning of the original content. The third task is content generation. LLMs can create a diverse array of content, ranging from emails and social media posts to blog articles and stories. The fourth task is question answering. In the context of question answering, LLMs either utilize the factual knowledge embedded in their pre-training corpus or the external contextual information provided in the prompt to generate sensible answers to questions or instructions. The fifth task is reasoning. Unlike common sense question answering, complex reasoning requires the understanding and application of evidence and logic to draw conclusions. This typically involves a sequential reasoning process based on factual knowledge, leading to the resolution of a posed question.

While the generation of executable software code is a significant capability of LLMs, it is not the primary focus in this context. This skill is more pertinent to product development and enhancement, where creating and refining software is essential. In contrast, technical service agents prioritize front-line problem solving and responding to user inquiries. Their role involves addressing immediate issues and providing clear, concise answers, rather than developing or improving software-based products.

### 3.2. Prototyping, Data and Technology Architecture

We use prototyping as a research methodology to validate three of the above cognitive tasks in a technical service setting: text correction, summarization and question answering. In design research, prototyping is used to concretize abstract concepts and validate their technical feasibility [23,24]. Therefore, prototyping is well suited to investigate theoretical approaches for the application of LLMs in manufacturing services that have been discussed in the previous literature.

In manufacturing, the proper operation and maintenance of machines is a knowledge-intensive task that requires technical domain expertise. Often, it also necessitates support from the machine manufacturers through technical service. Technical service agents address machine-related questions from operators and resolve incidents through direct communication with them.

We prototype the aforementioned tasks using real-life data on technical customer requests. These requests encompass IT-related topics such as software, hardware, and network issues. Although machine operation in manufacturing also involves non-IT-related issues, this dataset is well-suited for demonstrating technical service tasks because: 1. manufacturing increasingly incorporates IT topics, and 2. the data includes specialized technical domain topics which presents a major challenge in technical service for machines.

The data sets consist of the textual customer request and an exchange of messages by technical experts that describe one or more possible solutions. To enable manual validation and keep complexity low, we use a random selection of 15 customer requests in the prototypes.

The technical architecture described involves the integration of several advanced components to create a robust system for task completion and question answering. The implementation leverages the OpenAI API, utilizing the latest LLMs, specifically gpt-4-0125-preview and gpt-3.5-turbo-0125. These models are selected for their optimization in performing tasks and answering questions, representing the cutting edge in the field of LLMs at the time of writing [25].

A key component of the architecture is the use of Chroma DB, which serves as a vector database within the RAG framework. This database is complemented by OpenAI's text-embedding-3-small model, which is employed to generate embeddings for the text data. The overarching software framework used to integrate these components is LangChain.

The RAG design is further detailed with specific configurations and components. The LangChain docx2txt loader is used for document loading, and the text is split into chunks with a size of 1000 tokens and an overlap of 20 tokens to ensure context continuity. The retriever component utilizes LangChain's vector store with Chroma, configured with a parameter 'k' set to 2, indicating the number of nearest neighbors to retrieve.

The prompt template used in this architecture is shown below.

```
PROMPT_TEMPLATE = """
Answer the user questions in detail and explain all necessary solution steps. If the context does not contain any relevant information to answer the question, just say "I don't know":
<context>{context}</context>"""
```

For validation, we manually compare the LLM outputs with the messages and solutions generated by the human support agents and create quantitative quality metrics, which we discuss in detail below.

## 4. Results

### 4.1. Text Correction

LLMs can convert text from one language or language mode to another, as well as correct spelling and grammar errors. They achieve this by understanding the patterns and structures of different languages that they learn from the data they are trained with. This is crucial in technical service within the production machinery sector, where technology experts spend a significant amount of time communicating with machine operators.

For the PoC, we have written a reply email for each of the 15 customer incidents. Then we randomly added classic typos, especially letter twisters and missing letters. Table 2 shows the number of words, characters and spelling errors for the uncorrected reply emails. In the next step, we generated corrected reply mails with gpt-3.5-turbo-0125.

Table 2. Original Versus Corrected Replay Emails

| Incident | # Words Original | # Errors | # Words Final | # Errors Removed |
|---|---|---|---|---|
| 1 | 172 | 25 | 172 | 25 |
| 2 | 153 | 15 | 159 | 14 |
| 3 | 153 | 15 | 155 | 15 |
| 4 | 220 | 21 | 221 | 21 |
| 5 | 198 | 24 | 200 | 24 |
| 6 | 187 | 24 | 187 | 24 |
| 7 | 203 | 23 | 204 | 23 |
| 8 | 143 | 16 | 146 | 15 |
| 9 | 149 | 19 | 154 | 17 |
| 10 | 218 | 35 | 226 | 32 |
| 11 | 208 | 31 | 210 | 31 |
| 12 | 258 | 34 | 260 | 34 |
| 13 | 261 | 32 | 263 | 32 |
| 14 | 138 | 16 | 143 | 16 |
| 15 | 205 | 22 | 200 | 22 |

A manual analysis of the corrections as well as the quantitative key figures show that LLMs can generate reliable corrections without changing the wording and message content of the error-ridden email. Almost all typos were eliminated by this automated process.

*4.2. Text Summary*

Modern LLMs build on the transformer architecture and make use of a functionality known as the attention mechanism (Vaswani et al., 2017) This mechanism allows the model to consider the context of the use of each individual word. As a result, the model can understand the relationships between different sections of text and create a summary that accurately reflects the overall meaning of the original text.

In first-level support, it is important to quickly record the communication exchange that has already taken place on a customer request. Automatically generated text summaries could represent a high added value here.

To test this functionality, we had summaries of 100, 200 and 500 words created for each of the 15 data sets consisting of the customer request and the message exchange for solutions. We use gpt-4-0125-preview for summary generation.

For a systematic comparison of the summary with the source text, we calculate the cosine similarity (1=max, -1=min) of the two corresponding embedding vectors. Cosine similarity is a measure that measures the similarity between two vectors. The results are shown in Table 3. The cosine similarities are between .63 (minimum value) and .86 (maximum value). In addition, the cosine similarity increases with longer summaries.

In summary, it can be said that gpt-4-0125-preview is able to generate short summaries with a given word count relatively reliably. At the specification of 100 words, the text length average of the summaries produced was 100.33 words. Here, a comparative test with gpt-3.5-turbo-0125 resulted in significant text length deviations. However, with the text length specification of 500 words, gpt-4-0125-preview produced significantly shorter text (average: 268.4 words).

We conducted a manual analysis of the produced summaries to assess how well they captured all relevant information. For instance, one customer inquiry involved performing a software update. We manually extracted all pertinent solution steps from the customer request data (such as turning off the device, disconnecting the network, etc.) and evaluated the extent to which the summaries included these steps. The qualitative analysis of the summaries confirms the implication derived from the cosine similarities that the summaries reflect the essential contents of the solution discussion in a relatively robust manner.

Table 3. Cosine Similarities (CS) and Time Saved in Minutes (Min) for Summaries with 100, 200 and 500 Words

| Incident | CS 100 | CS 200 | CS 500 | Min 100 | Min 200 | Min 500 |
|---|---|---|---|---|---|---|
| Inc1 | 0.69 | 0.7 | 0.75 | 5.63 | 5.41 | 4.99 |
| Inc10 | 0.63 | 0.63 | 0.72 | 4.13 | 4 | 3.52 |
| Inc11 | 0.76 | 0.8 | 0.84 | 1.95 | 1.69 | 1.42 |
| Inc12 | 0.69 | 0.71 | 0.77 | 11.68 | 11.42 | 10.81 |
| Inc13 | 0.67 | 0.7 | 0.79 | 9.83 | 9.64 | 8.86 |
| Inc14 | 0.69 | 0.67 | 0.68 | 5.61 | 5.5 | 5.12 |
| Inc15 | 0.71 | 0.74 | 0.79 | 5.69 | 5.39 | 5.03 |
| Inc2 | 0.81 | 0.84 | 0.85 | 0.9 | 0.67 | 0.65 |
| Inc3 | 0.74 | 0.74 | 0.82 | 1.18 | 1.01 | 0.38 |
| Inc4 | 0.71 | 0.77 | 0.83 | 4.82 | 4.6 | 3.82 |
| Inc5 | 0.75 | 0.8 | 0.8 | 4.95 | 4.55 | 4.26 |
| Inc6 | 0.81 | 0.82 | 0.86 | 2.76 | 2.55 | 1.97 |
| Inc7 | 0.74 | 0.75 | 0.82 | 5.08 | 4.87 | 4.38 |
| Inc8 | 0.74 | 0.78 | 0.83 | 1.95 | 1.73 | 1.17 |
| Inc9 | 0.8 | 0.84 | 0.83 | 2.29 | 1.93 | 1.49 |

*4.3. Question Answering*

When answering questions, the LLM seeks and uses either the internal factual knowledge provided in the pre-training corpus or the external contextual data provided in the prompt to generate answers to questions or instructions. Technical service providers frequently handle customer inquiries regarding previously identified issues. To address these inquiries, historical data can be leveraged to develop solutions. In the production machinery sector, this historical data may include service incident databases, problem management and solution documentation, email correspondence, and machine documentation. Common customer inquiries in machine operations involve tasks such as machine configuration, troubleshooting outages, installation procedures, and maintenance activities. The analysis of such historical data can be automated with LLMs.

In a prototype, we generate 10 synthetic customer inquiries for every 10 customer inquiries, each addressing the same problem but using different formulations. First, we then use a vector search [19] to search the historical records for comparable customer requests. In vector search, the prompt is used as a search term and we specify how many similar chunks the search returns. An overview of the search results shows Table 4. If only one text chunk is returned per search, the 10 searches will return 100% text chunks that are relevant for problem solving because they are part of the correct data set. With three sections of text, the figure is still 87%.

Table 4. Proportion of Relevant Text Chunks of Vector Searches

| Number of Chunks in Vector Search | Average Proportion of Relevant Chunks |
|---|---|
| 1 | 100% |
| 2 | 95% |
| 3 | 87% |

In the next step, we embed vector search in a RAG architecture [19]. In the context field of the prompt, the result of the vector search is dynamically inserted. The synthetic customer request is then added to the prompt. We send the prompt assembled in this way to gpt-3.5-turbo-0125.

In order to validate the content of the LLM answers, we use vector search as well as qualitative comparisons. We enter the returned LLM answers into the vector search and extract the cosine distances (0=min, 2=max) to the nearest text sections of the 15 historical solution descriptions to the customer requests in columns Inc1 to Inc10. The results are depicted in Table 5.

Table 5. Cosine Distances of the 10 LLM Responses to the Nearest Text Chunks of the 15 Original Customer Requests

| Synthetic Request No: | 1 | 2 | 3 | 4 | 5 | 6 | 7 | 8 | 9 | 10 |
|---|---|---|---|---|---|---|---|---|---|---|
| Original Request: | | | | | | | | | | |
| Inc1 | 0.4 | 0.9 | 1.3 | 1 | 1.1 | 0.9 | 0.9 | 1.3 | 1.1 | 1 |
| Inc2 | 1.1 | 0.4 | 1.2 | 1.1 | 0.9 | 1.3 | 0.9 | 1.3 | 1.2 | 1.2 |
| Inc3 | 1.2 | 0.9 | 0.7 | 1.2 | 1.2 | 1.4 | 0.9 | 1.2 | 1.2 | 1.2 |
| Inc4 | 0.9 | 1 | 1.3 | 0.8 | 1.1 | 1 | 1 | 1.2 | 0.9 | 1 |
| Inc5 | 1.1 | 0.9 | 1.3 | 1.3 | 0.5 | 1.3 | 1 | 1.3 | 1.3 | 1.3 |
| Inc6 | 0.9 | 1.3 | 1.4 | 1.1 | 1.2 | 0.4 | 1 | 1.3 | 1.1 | 1 |
| Inc7 | 0.9 | 0.9 | 1.2 | 0.9 | 1.1 | 1 | 0.4 | 1.1 | 1 | 1 |
| Inc8 | 1.1 | 1.1 | 1.5 | 1.1 | 1.2 | 1.2 | 1 | 0.5 | 1.2 | 1.1 |
| Inc9 | 1.1 | 1 | 1.4 | 0.9 | 1.2 | 1.1 | 1.2 | 1 | 0.6 | 1 |
| Inc10 | 0.8 | 0.9 | 1.3 | 1 | 1.2 | 1.1 | 1 | 1.1 | 1 | 0.5 |
| Inc11 | 1 | 1 | 1.4 | 1.1 | 1.1 | 1.2 | 0.9 | 1.1 | 1 | 1.1 |
| Inc12 | 1 | 0.9 | 1.3 | 0.9 | 1.1 | 1.1 | 0.7 | 1 | 1.1 | 0.9 |
| Inc13 | 0.9 | 0.9 | 1.3 | 1 | 1.1 | 1.1 | 0.9 | 1.2 | 1.1 | 1.1 |
| Inc14 | 1.1 | 1 | 1.3 | 1.2 | 1.2 | 1.2 | 0.7 | 0.8 | 1.2 | 1.2 |
| Inc15 | 1.1 | 1.1 | 1.4 | 1.2 | 1.2 | 1.1 | 1 | 1.1 | 1.2 | 1.2 |

The results show that the cosine distances to the relevant text chunks of the original customer requests are minimal in each case (mean value 0.5). In addition, it can be seen that the distances to the most similar irrelevant text chunks are substantial (mean value 0.9). Inc4 is an exception. There, the cosine distance to the relevant section is high at 0.77 and the cosine distance to the nearest irrelevant section is relatively low at 0.9. However, a qualitative analysis of the synthetic customer request Inc4 showed that it was ambiguously formulated.

Overall, the qualitative and quantitative evaluations allow the conclusion that answering questions on the basis of historical data sets is also a promising use case for manufacturing service automation with LLMs.

## 5. Discussion

This research contributes to the theoretical understanding of integrating LLMs into technical service in the production machinery sector. It does so in several key areas.

First, this study identifies and validates three distinct cognitive tasks—translation and correction, summarization, and question answering—that LLMs can perform effectively in manufacturing contexts. By mapping these tasks to specific challenges in technical service for production machinery, the research extends the theoretical taxonomy of LLM capabilities and their practical applications in manufacturing.

Second, the integrating RAG into this study demonstrates the development of a domain-specific knowledge framework. By combining LLMs' generative abilities with external, domain-specific databases, the research presents a hybrid approach that improves the relevance and accuracy of outputs. This framework offers a scalable solution for incorporating continuous knowledge updates in dynamic industrial environments.

Finally, the study advances theoretical knowledge by prototyping technical service automation tasks. Prior has only begun to discuss the potential of LLMs for technical service in manufacturing research [8,9] .This research demonstrates the practical effectiveness of LLMs in automating text correction, summarization, and question answering tasks.

This research further provides practical insights into how LLMs can enhance efficiencies in manufacturing services through automation of key tasks. The study demonstrates practical applications of LLMs in text correction, summarization, and question answering, each of which delivers measurable improvements in operational workflows.

LLMs deliver substantial efficiency gains in summarizing customer inquiries and solution discussions. By generating concise summaries of lengthy communication exchanges, LLMs enable customer service agents to quickly grasp the context and history of an issue. The study shows that LLMs can produce summaries with high semantic similarity to the original text while saving substantial processing time. This automation reduces the cognitive load on agents, and accelerates case resolution.

The research showcases how LLMs streamline question answering by leveraging historical datasets to provide accurate, context-aware responses to customer queries. The integration of vector searches and RAG ensures that responses are grounded in relevant historical records, significantly improving the precision and relevance of answers. The prototypes demonstrated that LLMs could retrieve and synthesize appropriate solutions from historical records with high accuracy. This automation minimizes the time agents spend searching for solutions in large datasets, allowing them to focus on resolving complex, non-standard issues.

By automating these tasks, LLMs not only improve operational efficiency but also enhance the quality of customer service. The reductions in manual effort for error correction, information synthesis, and query resolution free up valuable time for support teams, enabling them to focus on strategic and creative problem-solving activities.

## 6. Conclusion

This study demonstrates how LLMs can enhance efficiency in manufacturing services through automated text correction, summarization, and question answering. The findings highlight significant time savings and improved accuracy, enabling customer support teams to work more effectively and focus on complex tasks.

However, the research has limitations, including the reliance on a small sample size data set for prototyping, which may not fully capture the variability of real-world scenarios. Additionally, challenges such as ambiguous inputs, knowledge hallucination, and ensuring seamless integration with human workflows remain to be further addressed.

Future research should explore the scalability of these solutions in diverse manufacturing environments and evaluate their long-term impact on operational performance. Investigating ways to mitigate challenges, such as improving prompt engineering and refining LLM training with domain-specific data, will further support the practical adoption of these technologies.